\documentclass[english,prd,twocolumn,superscriptaddress,nofootinbib,preprintnumbers]{revtex4}
\usepackage[latin1]{inputenc}
\usepackage{graphicx}
\usepackage{amssymb}
\usepackage{bm}
\usepackage{color}

\usepackage{amsfonts}
\usepackage{dcolumn}
\usepackage{hyperref}

\def\be{\begin{equation}}
\def\ee{\end{equation}}
\def\ba{\begin{eqnarray}}
\def\ea{\end{eqnarray}}
\def\bs{\begin{subequations}}
\def\es{\end{subequations}}

\usepackage{color}

\makeatother

\usepackage{babel}
\makeatother
\begin{document}

\title{Observational constraints on assisted k-inflation}

\author{Junko Ohashi}
\affiliation{Department of Physics, Faculty of Science, Tokyo University of Science, 
1-3, Kagurazaka, Shinjuku-ku, Tokyo 162-8601, Japan}

\author{Shinji Tsujikawa}
\affiliation{Department of Physics, Faculty of Science, Tokyo University of Science, 
1-3, Kagurazaka, Shinjuku-ku, Tokyo 162-8601, Japan}

\begin{abstract}

We study observational constraints on the assisted k-inflation models
in which multiple scalar fields join an attractor characterized by
an effective single field $\phi$.
This effective single-field system is described by the Lagrangian 
$P=Xg(Y)$, where $X$ is the kinetic energy of $\phi$, $\lambda$ is a constant, and 
$g$ is an arbitrary function in terms of $Y=X e^{\lambda \phi}$.
Our analysis covers a wide variety of k-inflation models such as 
dilatonic ghost condensate, Dirac-Born-Infeld (DBI) field, tachyon, 
as well as the canonical field with an exponential potential.
We place observational bounds on the parameters of each model from 
the WMAP 7yr data combined with 
Baryon Acoustic Oscillations (BAO) and the Hubble constant measurement.
Using the observational constraints of the equilateral non-Gaussianity 
parameter $f_{\rm NL}^{\rm equil}$, we further restrict
the allowed parameter space of dilatonic ghost condensate and DBI models.
We extend the analysis to more general models with several different 
choices of $g(Y)$ and show that the models such as 
$g(Y)=c_0+c_p Y^p$ ($p \ge 3$) are excluded by 
the joint data analysis of the scalar/tensor spectra and
primordial non-Gaussianities.

\end{abstract}

\date{\today}


\maketitle

\section{Introduction}

The cosmic acceleration in the early Universe--inflation-- has been 
the backbone of the high-energy cosmology over the past 
30 years. In addition to addressing the horizon and flatness problems 
plagued in Big Bang cosmology \cite{infpapers}, 
inflation generally predicts almost scale-invariant adiabatic 
density perturbations \cite{infper} (see \cite{review1,review2,review3} 
for reviews). This prediction is consistent with the observations of the 
Cosmic Microwave Background (CMB) temperature 
anisotropies measured by COBE \cite{COBE} 
and WMAP \cite{Komatsu:2010fb}.
It is possible to distinguish between a host of inflationary models
by comparing the theoretical prediction of the spectral index 
$n_{\rm s}$ of curvature perturbations and the tensor-to-scalar 
ratio $r$ with observations, but still the current observations
are not sufficient to identify the best model of inflation.

In the next few years, the measurement of CMB temperature 
anisotropies by the PLANCK satellite \cite{PLANCK} will bring 
more high-precision data. In addition to the possible reduction 
of the tensor-to-scalar ratio to the order of 0.01, 
the non-linear parameter $f_{\rm NL}$ of primordial scalar 
non-Gaussianities may be constrained by about one order of 
magnitude better than the bounds constrained by the WMAP group.
This can potentially provide further important information to discriminate 
between many inflation models.

The conventional single-field inflation driven by a canonical scalar field 
$\phi$ with a potential $V(\phi)$ predicts small primordial
non-Gaussianities with $f_{\rm NL}$ of the order of slow-roll
parameters \cite{Bartolo,Maldacena,Lyth} 
(see \cite{early} for early works).
However the kinetically driven inflation models 
(dubbed ``k-inflation'' \cite{kinflation}) 
described by the Lagrangian density $P(\phi, X)$, 
where $X$ is the field kinetic energy, 
can give rise to large non-Gaussianities with 
$|f_{\rm NL}| \gg 1$ \cite{Seery,Chen}.
This is related to the fact that for the Lagrangian 
including a non-linear kinetic term of $X$ the propagation 
speed $c_s$ is different from 1 (in the unit where the speed 
of light $c$ is $1$) \cite{Garriga,Hwang,DeFelice}.
Since the non-linear parameter is approximately given by 
$f_{\rm NL} \sim -1/c_s^2$, one has $|f_{\rm NL}| \gg 1$
for $c_s^2 \ll 1$.

In the models motivated by particle physics such as superstring and supergravity 
theories, there are many scalar fields that can be responsible for 
inflation \cite{review2,review3}.
In some cases, even if each field is unable to lead to cosmic acceleration, 
the presence of many fields allows a possibility for the realization of inflation
through the so-called assisted inflation mechanism \cite{Liddle}. 
In fact, multiple (canonical) scalar fields with exponential potentials 
$V_i(\phi_i)=c_i e^{-\lambda_i \phi_i}$ evolve to give dynamics 
matching a single field with the effective slope 
$\lambda=\left( \sum_{i=1} 1/\lambda_i^2 
\right)^{-1/2}$ \cite{Liddle}. 
Since $\lambda$ is smaller than the individual 
$\lambda_i$, the presence of multiple fields can lead to 
sufficient amount of inflation \cite{assistedpapers}.

If we take into account a barotropic perfect fluid (density $\rho_m$)
in addition to the canonical scalar field (density $\rho_{\phi}$) 
with the exponential potential $V(\phi)=c e^{-\lambda \phi}$, 
there exists a so-called scaling solution 
along which the ratio $\rho_{\phi}/\rho_m$ is 
constant \cite{Ferreira,CLW}.
In the presence of non-relativistic matter the scaling solution is unstable 
for $\lambda^2<3$, in which case
another scalar-field dominated solution is a stable attractor \cite{CLW}.
If $\lambda^2<2$, the latter can be used for inflation as well as dark energy.
If we extend the analysis to the models described by the general Lagrangian 
$P(\phi, X)$ then the condition for the existence of scaling solutions restricts
the form of the Lagrangian to be $P=Xg(Y)$, where 
$\lambda$ is a constant and $g$ is an arbitrary function in terms 
of $Y=Xe^{\lambda \phi}$ \cite{PT04,TS04}. 
Provided $\lambda^2<2\,\partial P/\partial X$ 
there exists a scalar-field dominated attractor 
that can be responsible for inflation \cite{Tsujikawa06,Amendola}. 
In fact this Lagrangian covers a wide class of inflationary models 
such as the canonical scalar field with the exponential potential
($g(Y)=1-c/Y$, i.e. $P=X-ce^{-\lambda \phi}$) \cite{exppapers} and 
the dilatonic ghost condensate model 
($g(Y)=-1+cY$, i.e. $P=-X+ce^{\lambda \phi}X^2$)
\cite{PT04} (see Refs.~\cite{Arkani,Arkani2} for the original 
ghost condensate model).

In the presence of multiple scalar fields it was shown that the Lagrangian 
$P=\sum_{i=1}^n X_i g(Y_i)$, where $g(Y_i)$ is an arbitrary function 
with respect to $Y_i=X_i e^{\lambda_i \phi_i}$, 
gives rise to assisted inflation \cite{Tsujikawa06,Ohashi},  
as it happens for the canonical field with 
the exponential potential.
In other words, in the regime where the solutions approach 
the assisted inflationary attractor, the system can be described by 
the effective single-field Lagrangian $P=X g(Y)$
with $Y=X e^{\lambda \phi}$ and the 
slope $\lambda=\left( \sum_{i=1}^{n} 1/\lambda_i^2 
\right)^{-1/2}$. While the scalar propagation speed is
different from 1 in those models, the scalar spectral index 
$n_{\rm s}$ and the tensor-to-scalar $r$ are written
in terms of the function $g(Y)$ and its derivatives $g'(Y)$, 
$g''(Y)$. By specifying the functional form of $g(Y)$, 
the observables $n_{\rm s}$ and $r$ as well as the 
equilateral non-Gaussianity parameter 
$f_{\rm NL}^{\rm equil}$ can be expressed by 
the single parameter $\lambda$ in the attractor regime.
This property is useful to place tight observational bounds
on those models.

In this paper we confront the assisted k-inflation scenario
described by the effective single-field Lagrangian 
$P=X g(X e^{\lambda \phi})$ with the recent CMB
observations by WMAP \cite{Komatsu:2010fb} combined 
with BAO \cite{BAO} and the Hubble constant 
measurement (HST) \cite{HST}.
We evaluate three observables $n_{\rm s}$, $r$, and 
$f_{\rm NL}^{\rm equil}$ without specifying the forms 
of $g(Y)$ and apply those results to concrete models
of inflation. We place observational constraints on a number of 
assisted inflation models such as 
(A) canonical field with the exponential potential,
(B) tachyon \cite{Sen}, (C) dilatonic ghost condensate, and 
(D) DBI field \cite{DBI}. 
Since the effect of the non-linear term in $X$ is important 
in the models (C) and (D), the primordial 
non-Gaussianity can reduce the parameter space constrained
by the information of $n_{\rm s}$ and $r$.

We shall also study other assisted inflation models such as
$g(Y)=c_0+\sum_{p \neq 0}c_p Y^p$ and the generalization of
the DBI model. Interestingly the observational bound from 
the equilateral non-Gaussianity parameter $f_{\rm NL}^{\rm equil}$
combined with $n_{\rm s}$ and $r$ can rule out some of those 
models. 

\section{Background dynamics in assisted k-inflation}
\label{dynamical}

We start with the single-field k-inflation models 
described by the action \cite{kinflation}
\begin{equation}
S=\int d^4x \sqrt{-g_M} \left[ \frac{R}{2} 
+ P(\phi,X) \right] \,,
\label{kinfaction}
\end{equation}
where $g_M$ is a determinant of the metric $g_{\mu \nu}$,
$R$ is a scalar curvature, $P$ is a general function in terms of 
the scalar field $\phi$ and 
the kinetic term $X=-g^{\mu\nu} \partial_{\mu}\phi\partial_{\nu}\phi/2$.
We use the unit $M_{\rm pl}=1$, where $M_{\rm pl}=(8\pi G)^{-1/2}$ is the
reduced Planck mass ($G$ is gravitational constant), but we restore
$M_{\rm pl}$ when the discussion becomes more transparent.

The pressure $P$ and the energy density $\rho$ of the field
$\phi$ are given, respectively, by
\begin{equation}
P=P(\phi, X)\,, \qquad \rho = 2XP_{,X}-P \,,
\end{equation}
where $P_{,X} \equiv \partial P/\partial X$.
We also define the equation of state $w_{\phi}$, as
$w_{\phi} \equiv P/\rho=P/(2XP_{,X}-P)$.
The cosmic acceleration can be realized under the condition 
$|2XP_{,X}| \ll |P|$, i.e. either (i) $X$ is small, 
or (ii) $P_{,X}$ is small. The case (i) corresponds to 
conventional slow-roll inflation driven by a field potential, 
whereas the case (ii) to kinetically driven inflation \cite{kinflation}.
One of the examples in the class (ii) is the ghost condensate
model \cite{Arkani,Arkani2} described by the Lagrangian 
$P=-X+X^2/M^4$, in which case inflation occurs 
around $X=M^4/2$.

In Refs.~\cite{PT04,TS04} it was shown that 
the condition for the existence of cosmological scaling 
solutions in the presence of non-relativistic matter 
restricts the Lagrangian of the form
\begin{equation}
P(\phi,X)=Xg (Y)\,,\qquad
Y \equiv Xe^{\lambda \phi}\,,
\label{singlelag}
\end{equation}
where $\lambda$ is a constant and $g$ is an arbitrary function 
in terms of $Y$.
This Lagrangian was derived by imposing that 
$\Omega_{\phi}/\Omega_m=$\,constant and $w_{\phi}=$\,constant 
in the scaling regime (where $\Omega_{\phi}$ and $\Omega_m$
are the density parameters of the scalar field and non-relativistic 
matter, respectively).

For the Lagrangian (\ref{singlelag}) there is another solution 
that can be responsible for the cosmic acceleration.
This corresponds to the fixed point 
with the equation of state \cite{Tsujikawa06}
\begin{equation}
w_{\phi}=-1+\frac{\lambda^2}{3P_{,X}}\,.
\end{equation}
The condition for the cosmic acceleration is $w_{\phi}<-1/3$, i.e. 
$\lambda^2<2P_{,X}$.
Since this point is stable for $\lambda^2<3P_{,X}$, 
the solutions approach it provided that inflation occurs.
Under the condition $\lambda^2<3P_{,X}$ the scaling solution is 
unstable \cite{Tsujikawa06}.

Let us consider the models with multiple scalar fields 
$\phi_i$ ($i=1,2,\cdots, n$) described by the action
\begin{equation}
S=\int d^4x \sqrt{-g_M} \left[ \frac{R}{2}
+ \sum_{i=1}^n X_i g (X_i e^{\lambda_i \phi_i}) \right] \,,
\label{multiaction}
\end{equation}
where $X_i=-g^{\mu\nu} \partial_{\mu}\phi_i \partial_{\nu} \phi_i/2$, 
$\lambda_i$'s are constants, and $g$ is an arbitrary function in terms of 
$Y_i= X_i e^{\lambda_i \phi_i}$.
Since we focus on inflation in the early Universe, 
we do not take into account other matter sources 
in the action (\ref{multiaction}).
In the flat Friedmann-Lema\^{i}tre-Robertson-Walker (FLRW) 
background with a scale factor $a(t)$, the 
equations of motion are
\begin{eqnarray}
& & 3H^2=\sum_{i=1}^n \rho_i\,,\\
& & 2\dot{H}=-\sum_{i=1}^n (P_i+\rho_i)\,,\\
& & \dot{\rho}_i+3H(\rho_i+P_i)=0\,,\qquad
(i=1, 2, \cdots, n),
\end{eqnarray}
where $H\equiv \dot{a}/a$ is the Hubble parameter 
(a dot denotes a derivative with respect to $t$), and 
\begin{equation}
P_i=X_i g (Y_i),\quad
\rho_i=X_i \left[ g(Y_i)+2Y_i g' (Y_i) \right].
\end{equation}
Here and in the following a prime represents 
a derivative of the corresponding quantities, e.g.,
$g'(Y_i)=dg/dY_i$.

In order to discuss the cosmological dynamics 
for the theories described by the action (\ref{multiaction})
we introduce the following quantities
\begin{equation}
x_i=\frac{\dot{\phi}_i}{\sqrt{6}H}\,,\qquad
y_i=\frac{e^{-\lambda_{i}\phi_i/2}}{\sqrt{3}H}\,.
\end{equation}
The differential equations for the variables $x_i$ and 
$y_i$ are given by 
\begin{eqnarray}
\hspace{-0.5cm}\frac{dx_i}{dN}&=&\frac{3x_{i}}{2}
\left[1+\sum_{i=1}^n g(Y_i)x_i^2-\frac{\sqrt{6}}{3}
\lambda_{i}x_{i}\right]+\frac{\sqrt{6}A(Y_i)}{2}  \nonumber \\
&& \times \left[\lambda_{i} 
\Omega_{\phi_i}-\sqrt{6}\{g(Y_i)+Y_ig'(Y_i)\}
x_{i} \right]\,, 
\label{auto1}\\
\hspace{-0.5cm}\frac{dy_i}{dN}&=&\frac{3y_i}{2}
\left[1+\sum_{i=1}^n g(Y_i)x_i^2-\frac{\sqrt{6}}{3}\lambda_{i}
x_{i} \right]\,,
\label{auto2}
\end{eqnarray}
where $N=\ln a$ is the number of e-foldings, and 
\begin{eqnarray}
A(Y_i) &=& \left[ g(Y_i)+5Y_ig'(Y_i)+2Y_i^2 
g'' (Y_i) \right]^{-1},\\
\Omega_{\phi_i} &=& x_i^2 \left[g(Y_i)+
2Y_i g'(Y_i) \right]\,.
\label{Omephi}
\end{eqnarray}

{}From Eqs.~(\ref{auto1}) and (\ref{auto2}) we find that 
the fixed point ($dx_i/dN=0$ and $dy_i/dN=0$) responsible 
for inflation ($y_i \neq 0$) satisfies
\begin{equation}
\label{lam1}    
\lambda_i x_i =
\frac{\sqrt{6}[g(Y_i)+Y_ig'(Y_i)]}
{g(Y_i)+2Y_ig'(Y_i)}=
\frac{\sqrt{6}}{2}
\left[1+\sum_{i=1}^n g(Y_i)x_i^2 \right]\,.
\end{equation}
Then the equation of state for each field, 
$w_{\phi_i}=g(Y_i)/[g(Y_i)+2Y_ig'(Y_{i})]$, reads
\begin{eqnarray}
\label{multieff}    
w_{\phi_i}=\sum_{i=1}^n g(Y_i) x_i^2=
-1+\frac{\sqrt{6}}{3}
\lambda_i x_i\,.
\end{eqnarray}
We require that Eq.~(\ref{lam1}) is satisfied for all 
$i=1, 2, \cdots, n$. Hence $\lambda_i x_i$'s 
are independent of $i$, i.e.
\begin{equation}
\label{lamx}
\lambda_1 x_1=\cdots=\lambda_i x_i=
\cdots =\lambda_n x_n \equiv \lambda x\,.
\end{equation}
This property also holds for $Y_i$ and $w_{\phi_i}$:
\begin{eqnarray}
\label{Yi}
& & Y_1=\cdots=Y_i=\cdots =Y_n \equiv Y\,,\\
& & w_{\phi_1}=\cdots=w_{\phi_{i}}=
\cdots =w_{\phi_{n}} \equiv w_{\phi}\,.
\end{eqnarray}
{}From Eq.~(\ref{lam1}) it follows that 
\begin{eqnarray}
\label{single1}    
\lambda x &=& \frac{\sqrt{6}[g(Y)+Yg'(Y)]}
{g(Y)+2Yg'(Y)} \\
&=&\frac{\sqrt{6}}{2}
\left[ 1+g(Y) x^2\lambda^2 
\sum_{i=1}^n \frac{1}{\lambda_i^2}\right]\,.
\label{single2} 
\end{eqnarray}

If we choose
\begin{eqnarray}
\label{efflambda}    
\frac{1}{\lambda^2}=
\sum_{i=1}^n \frac{1}{\lambda_i^2}\,,
\end{eqnarray}
then Eq.~(\ref{single2}) yields
\begin{eqnarray}
\label{single3}  
\lambda x=\frac{\sqrt{6}}{2}
\left[ 1+g(Y) x^2\right]\,.
\end{eqnarray}
This shows that, along the inflationary fixed point, the system 
effectively reduces to that of the single field with the Lagrangian 
$P=Xg (Y)$ with $Y=Xe^{\lambda \phi}$.
Since the sum of the density parameters 
$\Omega_{\phi_i}=x_i^2 \left[g(Y_i)+2Y_i g'(Y_i) \right]$
satisfies the relation $\sum_{i=1}^n \Omega_{\phi_i}=1$, 
we have
\begin{equation}
x^2 \left[ g(Y)+2Yg'(Y) \right]=1\,.
\label{Ometotal}
\end{equation}
{}From Eq.~(\ref{single1}) it then follows that 
$x=\lambda/(\sqrt{6}P_{,X})$, where we have 
used $P_{,X}=g(Y)+Yg'(Y)$.
The field equation of state (\ref{multieff}) is given by 
\begin{equation}
\label{weff}    
w_{\phi}=-1+\frac{\lambda^2}{3P_{,X}}\,.
\end{equation}

{}From Eq.~(\ref{efflambda}) we find that the effective slope squared 
$\lambda^2$ is smaller than $\lambda_i^2$ of each field. 
Even when the cosmic acceleration does not occur with a single field, 
it is possible to realize inflation in the presence of multiple fields.
The above discussion shows that assisted inflation occurs for the multi-field
k-inflation models described by the action (\ref{multiaction}).
In the regime where the solutions approach the assisted inflationary 
attractor satisfying the condition $\lambda^2<2P_{,X}$, the 
multi-field system reduces to that of the effective single field. 
In the following we shall study the effective single-field system  
described by the Lagrangian (\ref{singlelag}) with the 
slope $\lambda$ given in Eq.~(\ref{efflambda}).
As we have mentioned in Introduction, this
analysis covers a wide variety of 
assisted inflation models.

\section{Inflationary observables}
\label{observables}

It is possible to distinguish between a host of inflationary models 
by considering the spectra of primordial density 
perturbations generated during inflation.
For the calculations including primordial 
non-Gaussianities it is convenient to use the ADM 
metric \cite{Arnowitt} of the form 
\begin{eqnarray}
ds^{2} &=& -\left[ (1+\alpha)^{2}-a^{-2}(t) 
e^{-2{\cal R}}(\partial\psi)^{2}
\right]\,dt^{2}+2\partial_{i}\psi\, dt\, dx^{i} 
\nonumber \\
& &+a^2(t) (e^{2{\cal R}} \delta_{ij}
+h_{ij})dx^i dx^j\,,
\label{eq:metrica}
\end{eqnarray}
where $\alpha$, $\psi$, and ${\cal R}$ are scalar perturbations, 
and $h_{ij}$ are tensor perturbations.
We do not take into account vector perturbations because
they rapidly decay during inflation.

In the metric (\ref{eq:metrica}) we have gauged away a field 
$E$ that appears as a form $E_{,ij}$ inside the last parenthesis.
This fixes the spatial part of the gauge-transformation vector
$\xi^{\mu}$. We also choose the uniform-field gauge such that 
the inflaton fluctuation $\delta \phi$ vanishes ($\delta \phi=0$), 
which fixes the time component of $\xi^{\mu}$.

Integrating the action (\ref{kinfaction}) by parts for the
metric (\ref{eq:metrica}) and using the background 
equations of motion, the second-order action for the 
curvature perturbation can be written as \cite{Garriga}
\begin{equation}
S_2=\int dt\,d^3 x\,a^3\,Q \left[ 
\dot{\cal R}^2-\frac{c_s^2}{a^2} \partial^i {\cal R}
\partial_i {\cal R} \right]\,,
\label{secondaction}
\end{equation}
where $Q \equiv \epsilon/c_s^2$, and 
\begin{equation}
\epsilon  \equiv -\frac{\dot{H}}{H^2}\,,\qquad
c_s^2 \equiv \frac{P_{,X}}{P_{,X}+2XP_{,XX}}\,.
\label{csdef}
\end{equation}
The conditions for the avoidance of ghosts and Laplacian 
instabilities correspond to $Q>0$ and $c_s^2>0$, 
respectively, which are equivalent to 
\begin{equation}
\epsilon>0\quad {\rm and} \quad
c_s^2>0\,.
\label{nocon}
\end{equation}
For the Lagrangian including a non-linear term in 
$X$ (i.e. $P_{,XX} \neq 0)$,
the scalar propagation speed $c_s$ is different from 1.

The equation for the Fourier mode of ${\cal R}$
follows from the action (\ref{secondaction}).
For the modes deep inside the Hubble radius we choose 
the integration constants of the solution of ${\cal R}$
to recover the Bunch-Davies vacuum state. 
After the perturbations leave the Hubble radius 
($c_s k \lesssim aH$, where $k$ is a wave number) 
the curvature perturbation is frozen,
so that the scalar power spectrum is given by \cite{Garriga}
\begin{equation}
{\cal P}_{\rm s}=\frac{1}{8\pi^2 M_{\rm pl}^2}
\frac{H^2}{c_s \epsilon}\,,
\end{equation}
which is evaluated at $c_s k=aH$.
The scalar spectral index is 
\begin{equation}
n_{\rm s}-1 \equiv 
\frac{d \ln {\cal P}_{\rm s}}{d \ln k} \biggr|_{c_sk=aH}=
-2\epsilon-\eta-s\,,
\label{ncalR}
\end{equation}
where 
\begin{equation}
\eta \equiv \frac{\dot{\epsilon}}{H\epsilon}\,,\qquad
s \equiv \frac{\dot{c}_s}{Hc_s}\,.
\label{slow2}
\end{equation}
Here we have assumed that the field propagation speed 
slowly changes in time, such that $|s| \ll 1$.

For the theories described by the action (\ref{kinfaction}) 
the tensor perturbation $h_{ij}$ satisfies the same equation 
of motion as that for a massless scalar field. 
Taking into account two polarization states, 
the spectrum of $h_{ij}$ and its spectral index are given, 
respectively, by \cite{Garriga}
\begin{eqnarray}
& & {\cal P}_{\rm t}=\frac{2H^2}{\pi^2 M_{\rm pl}^2}\,,\\
& & n_{\rm t} \equiv 
\frac{d \ln {\cal P}_{{\rm t}}}{d \ln k} \biggr|_{k=aH}=
-2\epsilon\,.
\label{ncalT}
\end{eqnarray}
The tensor-to-scalar ratio is 
\begin{equation}
r \equiv \frac{{\cal P}_{\rm t}}{{\cal P}_{\rm s}}=
16c_s \epsilon=-8c_s n_{\rm t}\,.
\label{tsratio}
\end{equation}

The non-Gaussianity of the curvature perturbation is known 
by evaluating the vacuum expectation value of the three-point correlation function 
$\langle {\cal R} ({\bm k}_1) {\cal R} ({\bm k}_2) {\cal R} ({\bm k}_3)
\rangle$, where $ {\cal R} ({\bm k}_i) $ is the Fourier mode with 
a wave number ${\bm k}_i$ ($i=1,2,3$).
We write the bispectrum in the form 
$\langle {\cal R} ({\bm k}_1) {\cal R} ({\bm k}_2) {\cal R} ({\bm k}_3)
\rangle =(2\pi)^3 \delta^{(3)} ({\bm k}_1+{\bm k}_2+{\bm k}_3)
({\cal P}_{\rm s})^2 B (k_1,k_2,k_3)$, where $k_i=|{\bm k}_i|$.
In k-inflation one can take a factorizable shape
function $B=(2\pi)^4(9f_{\rm NL}/10)[-1/(k_1^3k_2^3)-
1/(k_1^3k_3^3)-1/(k_2^3k_3^3)-2/(k_1^2k_2^2k_3^2)
+1/(k_1k_2^2k_3^3)+(5~{\rm perm.})]$, where the permutations 
act on the last term in parenthesis \cite{Cremi,Koyamareview}.
For the equilateral triangles ($k_1=k_2=k_3$), 
the non-linear parameter is given by \cite{Seery,Chen,DeFelice}
\begin{eqnarray}
f_{\rm NL}^{\rm equil} &=& 
\frac{5}{81} \left( \frac{1}{c_s^2}-1
-\frac{2\mu}{\Sigma} \right)-\frac{35}{108}
\left( \frac{1}{c_s^2}-1 \right) \nonumber \\
& & +\frac{55}{36} \frac{\epsilon}{c_s^2}+
\frac{5}{12}\frac{\eta}{c_s^2}-\frac{85}{54}
\frac{s}{c_s^2}\label{kinf_fnl}\,,
\end{eqnarray}
where 
\begin{eqnarray}
\Sigma & \equiv& 
X P_{,X}+2X^2 P_{,XX}=H^2 \epsilon/c_s^2 \,, \\
\mu & \equiv& X^2 P_{,XX}+2X^3 P_{,XXX}/3 \nonumber \\
&=& \frac{\Sigma}{6} \left( \frac{1}{c_s^2}-1+
\frac23 \frac{\epsilon}{\epsilon_X} \frac{s}{c_s^2}
\right) \,,
\label{mudef}
\end{eqnarray}
and $\epsilon_X \equiv -(\dot{X}/H^2)(\partial H/\partial X)$.
In the second line of Eq.~(\ref{mudef}) we have used 
$\dot{X}=-6Hc_s^2 X \epsilon_X/\epsilon$, which
follows from the background equation of the field $\phi$ \cite{Seery}.
Our sign convention of $f_{\rm NL}^{\rm equil}$ coincides with that 
in the WMAP 7yr paper \cite{Komatsu:2010fb}.
The observational bound on the equilateral non-linear parameter 
constrained by the WMAP 7yr data is 
\begin{equation}
-214<f_{\rm NL}^{\rm equil}<266 \qquad
(95\,\%\,{\rm CL}).
\label{fnlbound}
\end{equation}

Let us consider the case in which the multiple fields
join the effective single-field attractor characterized by 
the conditions (\ref{single1})-(\ref{Ometotal}).
{}From Eqs.~(\ref{single3}) and (\ref{Ometotal}) we obtain
\begin{equation}
\lambda^2=\frac{6 \, [g(Y)+Yg'(Y)]^2}
{g(Y)+2Yg'(Y)}\,.
\label{lamcon}
\end{equation}
By choosing a specific function $g(Y)$ and solving 
Eq.~(\ref{lamcon}), we can determine $Y$ 
in terms of $\lambda$ (i.e. $Y$ is constant).
The slow-roll parameter $\epsilon$ and 
the scalar propagation speed squared $c_s^2$ are 
\begin{eqnarray}
\epsilon &=& \frac{3\,[g(Y)+Yg'(Y)]}{g(Y)+2Yg'(Y)} \,, 
\label{epsilon} \\
c_s^2 &=& \frac{g(Y)+Yg'(Y)}{g(Y)+5Yg'(Y)+2Y^2g''(Y)} \,, 
\label{cs2}
\end{eqnarray}
which are functions of $Y$ only.
Then one has $\epsilon=$\,constant and 
$c_s^2=$\,constant on the inflationary attractor, 
thereby leading to $\eta=0$, $s=0$, and 
$\mu/\Sigma=(1/c_s^2-1)/6$.
{}From Eqs.~(\ref{ncalR}), (\ref{ncalT}), (\ref{tsratio}),
and (\ref{kinf_fnl})
the three inflationary observables reduce to 
\begin{eqnarray}
n_{\rm s}-1 &=& -2\epsilon=n_{\rm t}\,,
\label{obser1}\\
r &=& 16c_s \epsilon=8c_s (1-n_{\rm s})\,,
\label{obser2}\\
f_{\rm NL}^{\rm equil}&=&
-\frac{275}{972} \left( 
\frac{1}{c_s^2} -1 \right)+\frac{55}{36} 
\frac{\epsilon}{c_s^2}\,,
\label{obser3}
\end{eqnarray}
where $\epsilon$ and $c_s^2$ are given in 
Eqs.~(\ref{epsilon}) and (\ref{cs2}).
Since $\epsilon$ is constant along the inflationary attractor, 
there are no runnings for scalar and tensor perturbations.

Since $Y$ is known in terms of $\lambda$ for given $g(Y)$, 
all the observables in Eqs.~(\ref{obser1})-(\ref{obser3}) can be 
expressed by $\lambda$ (or $\epsilon$).
Observationally one can place the bounds on the parameter $\lambda$
for each model. In the following we shall proceed to the observational
constraints on assisted k-inflation models.

\section{Observational constraints on four models of 
assisted inflation}
\label{constraint}

In this section we study the observational constraints 
on a number of assisted inflation models by choosing 
specific forms of $g(Y)$.
These include 
(A) canonical field with an exponential potential [$g(Y)=1-c/Y$],
(B) tachyon [$g(Y)=-c\sqrt{1-2Y}/Y$], 
(C) dilatonic ghost condensate [$g(Y)=-1+cY$], and
(D) DBI field [$g(Y)=-\sqrt{1-2Y}/Y-c/Y$], 
where $c$ is constant.

In the model (A) one has $c_s^2=1$, so that the non-Gaussianity 
is small enough ($f_{\rm NL}^{\rm equil}=55 \epsilon/36 \ll 1$) 
to satisfy the observational bound (\ref{fnlbound}).
We can constrain either $\lambda$ or $\epsilon$  
by carrying out the CMB likelihood analysis with respect to 
$n_{\rm s}$, $r$, and $n_{\rm t}$.
In the tachyon model (B) the scalar propagation speed $c_s$
does not equal to 1, but the difference from 1 is required 
to be small.
Hence the situation is similar to that in the model (A).

For the models (C) and (D) $c_s$ can be 
much smaller than 1, while satisfying the condition $\epsilon \ll 1$.
In such cases it is possible to place tight bounds on 
the models from the primordial non-Gaussianities
in addition to those coming from $n_{\rm s}$, $r$, 
and $n_{\rm t}$.

\subsection{Canonical field with an exponential potential}

The canonical field with the exponential potential described 
by the Lagrangian $P=X-c\,e^{-\lambda \phi}$
corresponds to the choice
\begin{equation}
g(Y)=1-c/Y\,.
\label{model1}
\end{equation}
In this case one has $c/Y=6/\lambda^2-1$, 
$\epsilon=\lambda^2/2$, and $c_s^2=1$.
Inflation occurs for $\lambda^2 \ll 1$, i.e.
$X \ll ce^{-\lambda \phi}$.
The inflationary observables are
\begin{eqnarray}
n_{\rm s}-1 &=& n_{\rm t}=-\lambda^2\,, \label{nsexp}\\
r &=& 8\lambda^2\,, \label{rexp}\\
f_{\rm NL}^{\rm equil} &= & 55\lambda^2/72\,. \label{fnlexp}
\end{eqnarray}

\begin{figure}
\includegraphics[height=3.0in,width=3.0in]{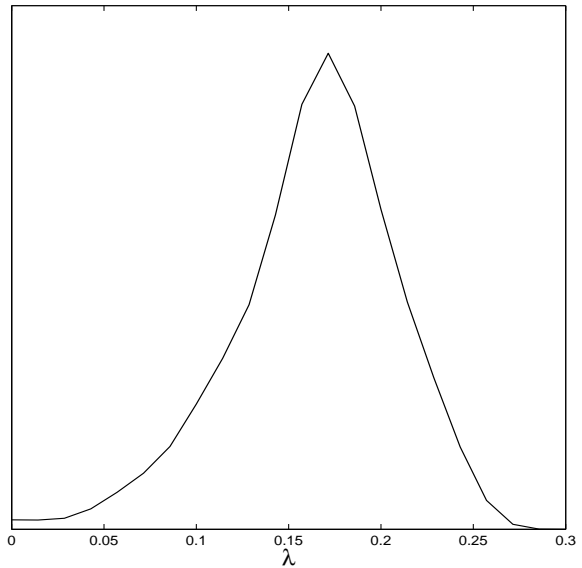}
\caption{\label{fig1}
One-dimensional marginalized probability distribution of the parameter $\lambda$ 
for the model $P=X-ce^{-\lambda \phi}$ constrained by the 
joint data analysis of WMAP 7yr, BAO, and HST.
We use the theoretical expression of $n_{\rm s}$, $r$, and $n_{\rm t}$ 
given in Eqs.~(\ref{nsexp}) and (\ref{rexp}).}
\end{figure}

Using the Cosmological Monte Carlo (CosmoMC) code \cite{cosmomc},
we carry out the likelihood analysis with the WMAP\,7yr data
combined with BAO and HST.
As we show in Fig.~\ref{fig1}, the likelihood analysis in terms of 
$n_{\rm s}$, $n_{\rm t}$, and $r$ gives the following bound
\begin{equation}
0.086<\lambda<0.228\qquad (95 \%~{\rm CL}).
\label{lamcan}
\end{equation}
The Harrison-Zel'dovich (HZ) spectrum ($n_{\rm s}=1$ and $r=0$)
is disfavored from the data.
Under the bound (\ref{lamcan}) one has $f_{\rm NL}^{\rm equil} \ll 1$, 
such that the non-Gaussianity constraint (\ref{fnlbound}) is satisfied.

%
\subsection{Tachyon}

A tachyon field $\varphi$ with a potential $V(\varphi)$ corresponds to 
the Lagrangian $P=-V(\varphi) \sqrt{1-2\tilde{X}}$, where 
$\tilde{X} \equiv -g^{\mu \nu} \partial_{\mu} \varphi 
\partial_{\nu} \varphi/2$ \cite{Sen}.
Choosing the function 
\begin{equation}
g(Y)=-c\sqrt{1-2Y}/Y\,,
\label{model2}
\end{equation}
where $Y=Xe^{\lambda \phi}=\tilde{X}$, 
one can show that the Lagrangian $P=Xg(Y)$ reduces to the form 
$P=-4c/(\lambda^2 \varphi^2) \sqrt{1-2\tilde{X}}$.
Hence the tachyon potential $V(\varphi) \propto \varphi^{-2}$
leads to assisted inflation. The cosmological dynamics 
in the presence of the inverse power-law tachyon potential 
have been discussed in Refs.~\cite{powerlawtachyon}.

For the choice (\ref{model2}) it follows that 
$\epsilon=3Y$ and $c_s^2=1-2Y$, where $Y$ is related 
to $\lambda$ via $\lambda^2=6cY/\sqrt{1-2Y}$.
The inflationary observables are 
\begin{eqnarray}
n_{\rm s}-1 &=& n_{\rm t}=-6Y\,,\\
r &=& 48 Y\sqrt{1-2Y}\,, \\
f_{\rm NL}^{\rm equil} &=&
\frac{3905}{972} \frac{Y}{1-2Y}\,.
\end{eqnarray}
Since we require $Y \ll 1$ to realize the nearly scale-invariant 
scalar spectrum, the non-Gaussianity is suppressed 
to be small ($f_{\rm NL}^{\rm equil} \ll 1$).
The relation between $r$ and $n_{\rm s}$ is given by 
$r=8(1-n_{\rm s})\sqrt{1-2Y} \simeq 8(1-n_{\rm s})$,
which, in the limit that $Y\to 0$, is the same as that for 
the canonical field with the exponential potential.
This property comes from the fact that tachyon inflation is 
driven by the potential energy rather than the field kinetic energy.
The joint CMB likelihood analysis combined with BAO and 
HST gives the bound 
\begin{equation}
1.7\times 10^{-3}<Y<7.7\times 10^{-3}
\qquad (95 \%~{\rm CL}).
\label{Ycon}
\end{equation}
Then $c_s^2=1-2Y$ is indeed close to 1.

\subsection{Dilatonic ghost condensate}

The dilatonic ghost condensate model is described by
the Lagrangian $P=-X+ce^{\lambda \phi} X^2$, i.e.
\begin{equation}
g(Y)=-1+cY\,.
\label{model3}
\end{equation}
In this case we have 
\begin{equation}
\epsilon=\frac{3(2cY-1)}{3cY-1}\,,\qquad
c_s^2=\frac{2cY-1}{6cY-1}\,,
\label{dilaep}
\end{equation}
where $cY$ is known by solving Eq.~(\ref{lamcon}), i.e.
\begin{equation}
2cY-1=f(\lambda)\,,\quad
f(\lambda) \equiv \frac{1}{8} \left[
\lambda^2+\sqrt{\lambda^4+
\frac{16}{3}\lambda^2} \right]\,.
\label{cY}
\end{equation}
In Eq.~(\ref{cY}) we have chosen the solution with $cY>1/2$
to avoid the appearance of ghosts \cite{PT04}.
The inflationary observables are given by 
\begin{eqnarray}
\hspace{-0.7cm}& & 
n_{\rm s}-1=n_{\rm t}=-\frac{\lambda^2}{f(\lambda)}\,,\label{nsdila}\\
\hspace{-0.7cm}& & r=\frac{8\lambda^2}{\sqrt{f(\lambda)[3f(\lambda)+2]}}\,, 
\label{rdila}\\
\hspace{-0.7cm}& & f_{\rm NL}^{\rm equil}=-\frac{275}{486} \left[
1+\frac{1}{f(\lambda)} \right]+\frac{55}{72} 
\frac{\lambda^2}{f(\lambda)} \left[ 3+
\frac{2}{f(\lambda)} \right].
\label{fnldila}
\end{eqnarray}
In the limit that $\lambda \to 0$ one has 
$f(\lambda) \simeq \lambda/2\sqrt{3} \to 0$ 
and hence $f_{\rm NL}^{\rm equil} \to -\infty$.
Using the WMAP 7yr bound $f_{\rm NL}^{\rm equil}>-214$, 
we obtain the constraint $\lambda>8.4 \times 10^{-3}$ (95 \%~{\rm CL}).

\begin{figure}
\includegraphics[height=3.0in,width=3.0in]{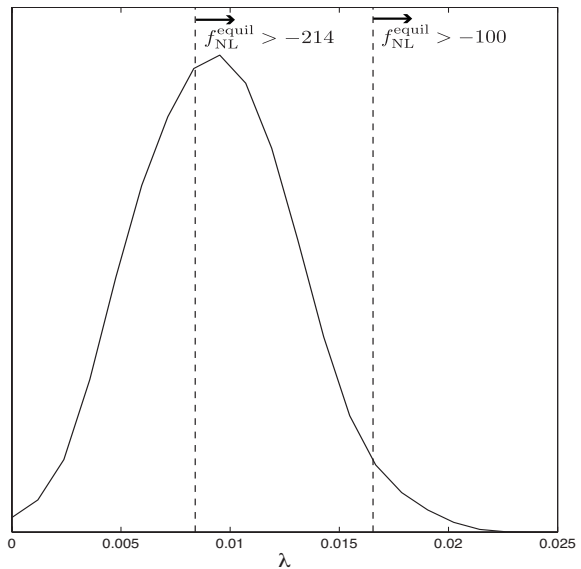}
\caption{\label{fig2}
One-dimensional marginalized probability distribution of the parameter $\lambda$ 
in the dilatonic ghost condensate model constrained by the 
joint data analysis of WMAP 7yr, BAO, and HST.
We also show the bound on $\lambda$ coming from the WMAP 7yr 
constraint of the equilateral non-Gaussianity parameter, $f_{\rm NL}^{\rm equil}>-214$, 
as well as the bound on $\lambda$ corresponding to the constraint 
$f_{\rm NL}^{\rm equil}>-100$. }
\end{figure}

In the region $\lambda^2 \ll 1$ one has $f(\lambda) \simeq \sqrt{3} \lambda/6$, 
$n_{\rm s} \simeq 1-2\sqrt{3} \lambda$, and 
$r \simeq 8 \cdot 3^{1/4} \lambda^{3/2}$, 
which give the relation $r \simeq (2\sqrt{6}/3)(1-n_{\rm s})^{3/2}$.
In this model the tensor-to-scalar ratio is smaller than the order of 0.1, so 
that the allowed region of $\lambda$ is mainly determined by $n_{\rm s}$. 
The CMB likelihood analysis in terms of $n_{\rm s}$, $n_{\rm t}$, $r$
shows that $\lambda$ is constrained to be 
$4.0\times10^{-3}<\lambda<1.5\times 10^{-2}$ 
(95 \%~{\rm CL}), see Fig.~\ref{fig2}.
Combining this with the non-Gaussianity constraint,
it follows that 
\begin{equation}
8.4 \times 10^{-3}<\lambda<1.5\times 10^{-2}
\qquad (95 \%~{\rm CL}).
\label{ghostcon}
\end{equation}

If the future observations constrain the non-Gaussianity parameter
at the level $f_{\rm NL}^{\rm equil}>-100$, it will be possible to 
exclude the dilatonic ghost condensate model (see Fig.~\ref{fig2}).
Moreover the precise measurement of the scalar index $n_{\rm s}$
can reduce the allowed range of $\lambda$ further.

%
\subsection{DBI field}

The DBI field $\phi$ is characterized by the Lagrangian 
\begin{equation}
P=-f(\phi)^{-1} \sqrt{1-2f(\phi)X}+f(\phi)^{-1}-V(\phi)\,,
\label{DBIlag}
\end{equation}
where $f(\phi)$ and $V(\phi)$ are functions of $\phi$.
If we choose
\begin{equation}
g(Y)=-\sqrt{1-2Y}/Y-c/Y\,,
\label{model4}
\end{equation}
the Lagrangian $P=Xg(Y)$ reduces to (\ref{DBIlag})
with $f(\phi)=e^{\lambda \phi}$ and 
$V(\phi)=(c+1)e^{-\lambda \phi}$.
Hence the DBI field with the exponential potential 
$V(\phi)=(c+1)e^{-\lambda \phi}$ leads to 
assisted inflation.

For the function (\ref{model4}) it follows that 
\begin{equation}
\epsilon=\frac{3Y}{c\sqrt{1-2Y}+1}\,,\quad
c_s^2=1-2Y\,.
\label{epDBI}
\end{equation}
If $c \lesssim 1$, one has $\epsilon \ll 1$ and $c_s^2 \simeq 1$
for $Y \ll 1$. This case is similar to tachyon inflation in which 
cosmic acceleration is driven by the field potential.
One can also realize $\epsilon \ll 1$ 
under the following condition 
\begin{equation}
c \sqrt{1-2Y} \gg 1\,.
\label{ccon}
\end{equation}
If $c \gg 1$, then it is possible to satisfy (\ref{ccon}) 
even for the values of $Y$ close to 1/2.
In fact this is the ultra-relativistic regime of the DBI inflation 
in which the $\gamma$ factor $\gamma=1/\sqrt{1-f(\phi) \dot{\phi}^2}$
is much larger than 1.
Even in this ``fast-roll'' regime the presence of the 
potential is important to satisfy the condition (\ref{ccon}).

\begin{figure}
\includegraphics[height=3.0in,width=3.0in]{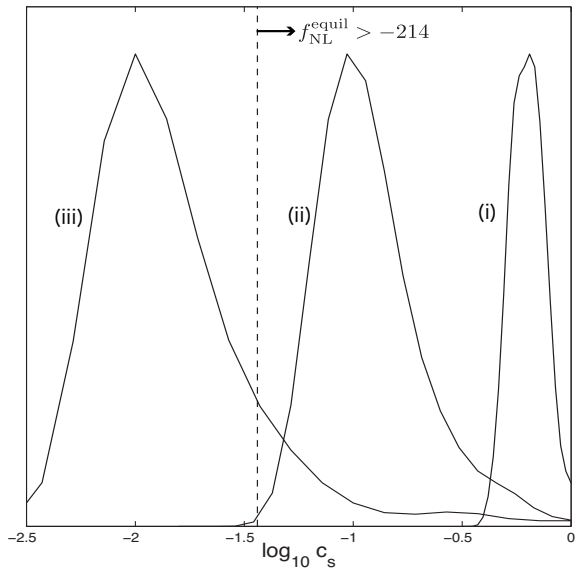}
\caption{\label{fig3}
One-dimensional marginalized probability distribution of 
the field propagation speed $c_s$ (with the logarithmic scale)
in the DBI model constrained by the observational data of 
WMAP 7yr, BAO, and HST. The three solid lines correspond to 
the cases: (i) $c=10^2$, (ii) $c=10^3$, and (iii) $c=10^4$.
We also show the bound derived from the non-Gaussianity 
constraint $f_{\rm NL}^{\rm equil}>-214$ in the limit 
$c c_s \gg 1$.}
\end{figure}

The inflationary observables are
\begin{eqnarray}
n_{\rm s}-1 &=& n_{\rm t}=-\frac{3(1-c_s^2)}{c\,c_s+1}\,,
\label{nsDBI}\\
r &=& \frac{24c_s(1-c_s^2)}{c\,c_s+1}\,, 
\label{rDBI}\\
f_{\rm NL}^{\rm equil} &=& 
-\frac{55}{1944}\frac{(10c\,c_s-71)}{(c\,c_s+1)}
\left( \frac{1}{c_s^2}-1 \right)\,, 
\label{fnlDBI}
\end{eqnarray}
where $c_s=\sqrt{1-2Y}$.
These observables depend not only on $c_s$ (or $Y$) but on 
the coefficient $c$ associated with the field potential.
For larger $c$ it is possible to satisfy the observational 
constraints of $n_{\rm s}$, $r$, and $n_{\rm t}$ with 
smaller $c_s$, because the denominators of 
Eqs.~(\ref{nsDBI}) and (\ref{rDBI}) get larger.
In fact, Fig.~\ref{fig3} shows that, for larger $c$, the one-dimensional 
marginalized probability distribution of $\lambda$ tends to 
shift to the regions of smaller $c_s$.
In Fig.~\ref{fig3} the propagation speed $c_s$ close to 1
is not favored because $n_{\rm s}$ and $r$ are close to the 
HZ spectrum.
The models with very small $c_s$ are also disfavored
because of the large deviation from the HZ spectrum.

\begin{figure}
\includegraphics[height=3.0in,width=3.5in]{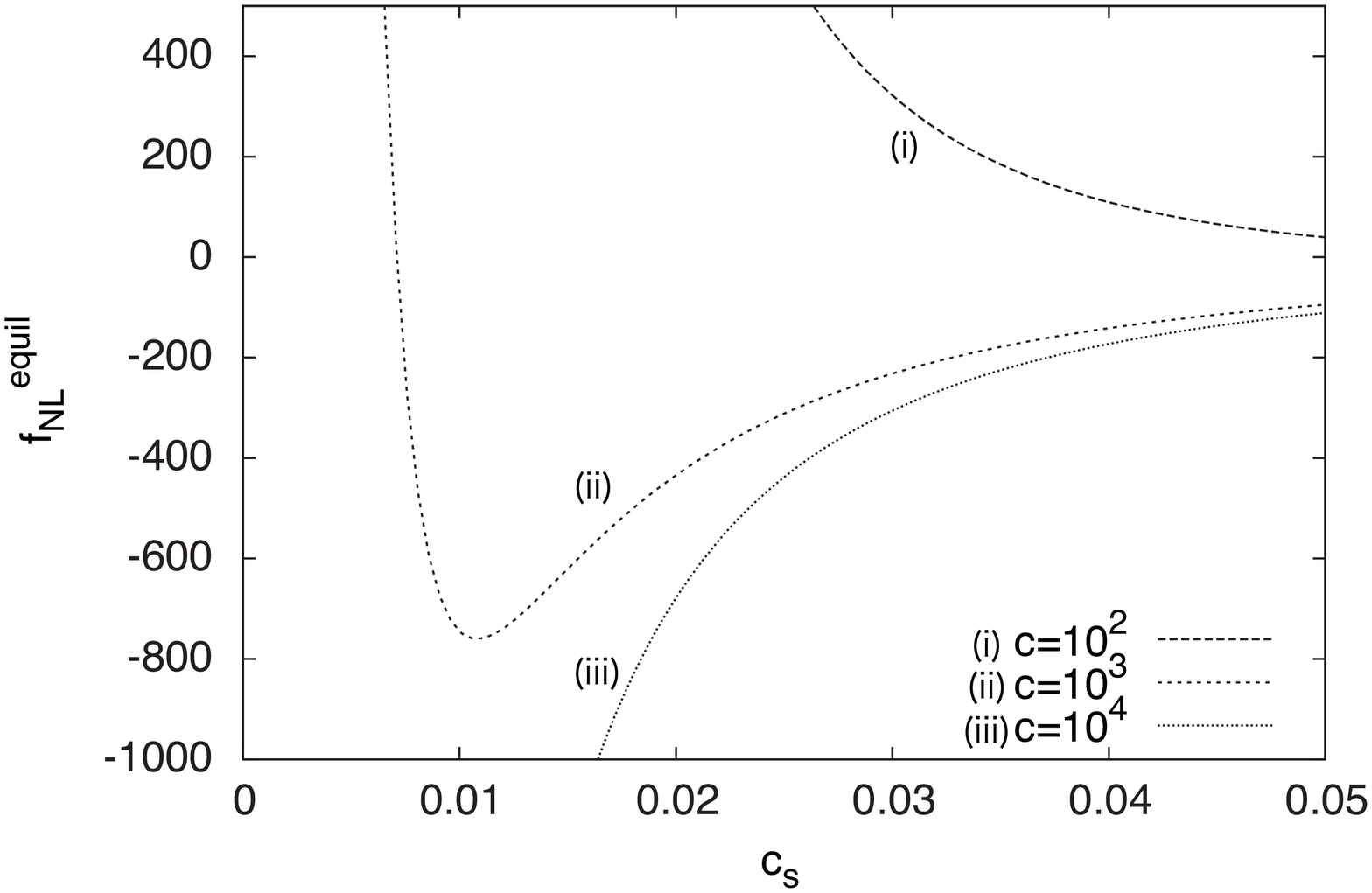}
\caption{\label{fig4}
The equilateral non-Gaussianity parameter $f_{\rm NL}^{\rm equil}$ versus 
the scalar propagation speed $c_s$ in the DBI model for 
(i) $c=10^2$, (ii) $c=10^3$, and (iii) $c=10^4$.
For $c=10^2$ the scalar propagation speed is constrained 
by the WMAP 7yr upper bound $f_{\rm NL}^{\rm equil}<266$, 
whereas for $c=10^3$ and $c=10^4$ it is constrained 
by the lower bound $f_{\rm NL}^{\rm equil}>-214$.}
\end{figure}

In Fig.~\ref{fig4} we plot the non-Gaussianity parameter $f_{\rm NL}^{\rm equil}$ 
given in Eq.~(\ref{fnlDBI}) versus the scalar propagation speed $c_s$
for three different values of $c$.
For $c=10^2$ we obtain the bound $c_s>3.2 \times 10^{-2}$ from 
the WMAP 7yr upper limit $f_{\rm NL}^{\rm equil}<266$.
On the other hand, the WMAP 7yr lower limit 
$f_{\rm NL}^{\rm equil}>-214$ gives the bounds
$c_s>3.1 \times 10^{-2}$ and $c_s>3.6\times 10^{-2}$
for $c=10^3$ and $c=10^4$, respectively.

As we see in Fig.~\ref{fig3}, the CMB likelihood analysis 
in terms of $n_{\rm s}$, $r$, and $n_{\rm t}$
places the constraints on $c_s$, as $0.48<c_s<0.84$ (95 \%~{\rm CL})
for $c=10^2$ and $0.06<c_s<0.35$ (95 \%~{\rm CL}) for $c=10^3$.
If $c \lesssim 10^3$ the non-Gaussianity does not provide additional 
constraints on $c_s$ to those derived by the likelihood analysis
in Fig.~\ref{fig3}.
If $c \gtrsim 10^3$ the non-Gaussianity plays an important 
role to restrict the allowed parameter space of $c_s$ further.
In particular, for $c=10^4$, there are almost no allowed regions
to satisfy all the observational constraints (see Fig.~\ref{fig3}).
Hence the models with $c \gtrsim 10^4$
are excluded by the analysis including non-Gaussianities.

\section{More general models}

So far we have studied the observational constraints on four 
assisted inflation models.
Among them the dilatonic ghost condensate and the DBI models 
can be tightly constrained by taking into account the bound
coming from the primordial non-Gaussianity.
This is associated with the fact that both $\epsilon$ and $c_s^2$
can be much smaller than 1 in those models.
In this section we shall extend the analysis to more general 
functions of $g(Y)$.

In the dilatonic ghost condensate model the numerators
of $\epsilon$ and $c_s^2$ in Eq.~(\ref{dilaep}) vanish 
at $cY=1/2$, whereas the denominators of them are 
non-zero finite values.
In the DBI model the numerator of $\epsilon$ in Eq.~(\ref{epDBI})
does not vanish in the ultra-relativistic regime 
($Y \approx 1/2$), whereas $c_s^2 \ll 1$.
In the DBI case it is possible to have $\epsilon \ll 1$ 
as long as the denominator of $\epsilon$ is much larger 
than the numerator of it [which is satisfied 
under the condition (\ref{ccon})].
Since these models are qualitatively different,
we classify the assisted k-inflation models into 
two classes in the following discussion.

\subsection{Class (i)}

Let us first study the models in which inflation occurs 
around $Y=Y_0$, where $Y_0$ satisfies
\begin{equation}
g(Y_0)+Y_0 g'(Y_0)=0\,.
\label{gY0}
\end{equation}
As in the case of the dilatonic ghost condensate, 
we consider the models in which the numerators of 
$\epsilon$ and $c_s^2$ in Eqs.~(\ref{epsilon}) and (\ref{cs2}) 
vanish, whereas the denominators are non-zero. 
Since $Y=Y_0$ corresponds to the exact 
de Sitter solution, we perform the linear expansion of 
the variables $\epsilon (Y)$ and $c_s^2 (Y)$ by setting 
$Y=Y_0+\delta Y$ with $|\delta Y/Y_0| \ll 1$.
It then follows that 
\begin{eqnarray}
\epsilon (Y) &\simeq& \epsilon' (Y_0)\,\delta Y
=\frac{6}{Y_0} \left[ 1+\frac{Y_0 g'' (Y_0)}{2g'(Y_0)} \right]
\,\delta Y\,,\\
c_s^2 (Y) &\simeq& {c_s^2}' (Y_0)\,\delta Y
=\frac{1}{2Y_0}\,\delta Y\,.
\label{expand}
\end{eqnarray}
This shows that the ratio $c_s^2/\epsilon$ is 
approximately constant in the regime $|\delta Y/Y_0| \ll 1$:
\begin{equation}
\frac{c_s^2}{\epsilon} \simeq \frac{1}{12}
\left[ 1 +\frac{Y_0 g'' (Y_0)}{2g'(Y_0)} \right]^{-1}\,.
\label{ratioapp}
\end{equation}

Expanding Eq.~(\ref{lamcon}) at $Y=Y_0$, we have
\begin{equation}
(\delta Y)^2=\frac{Y_0 g'(Y_0)}{6[2g'(Y_0)+
Y_0 g''(Y_0)]^2}\lambda^2\,.
\end{equation}
As long as $g'(Y_0)>0$ there exists a solution 
with $\delta Y>0$.
The conditions (\ref{nocon}) for the avoidance of 
ghosts and Laplacian instabilities translate into
\begin{eqnarray}
\label{dgy1} g'(Y_0) &>&0\,,\\
\label{dgy2} Y_0g''(Y_0) &>&-2g'(Y_0)\,.
\end{eqnarray}

In the ghost condensate model described by the function 
$g(Y)=-1+cY$ the second derivative $g''(Y)$ automatically 
vanishes, which gives $c_s^2/\epsilon=1/12$.
In this model the variable $\lambda$ is observationally 
bounded as $\lambda<1.5 \times 10^{-2}$ (95 \% CL), 
in which case $\delta Y/Y_0=\lambda/(2\sqrt{3})
<4.3 \times 10^{-3}$.
Hence it is a good approximation to use the linear expansion 
given above. In fact we have carried out the CMB likelihood 
analysis by employing the relation $c_s^2/\epsilon=1/12$
and confirmed that the observational bound on $\lambda$
is very similar to that given in Eq.~(\ref{ghostcon}).

\begin{figure}
\includegraphics[height=3.0in,width=3.5in]{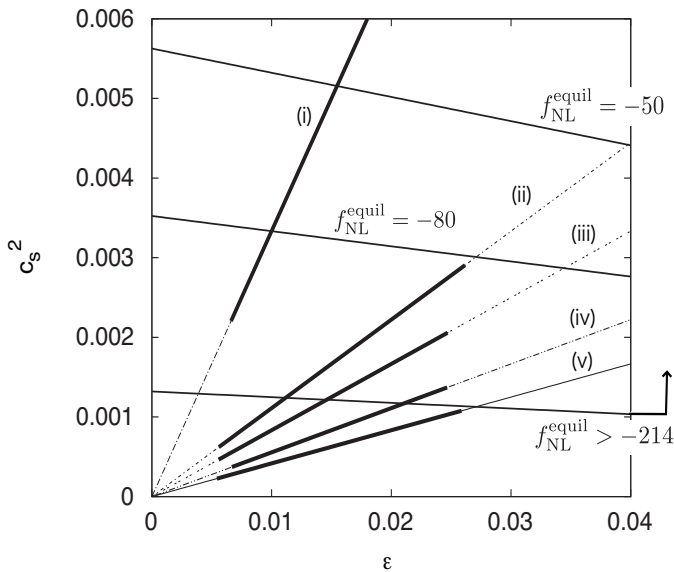}
\caption{\label{fig5}
The thin lines show the relations between $\epsilon$ and $c_s^2$ 
for the models $g(Y)=c_0+c_p Y^p$ with 
(i) $p=-0.5$, (ii) $p=0.5$, (iii) $p=1$, (iv) $p=2$, and 
(v) $p=3$, in the regime $\epsilon \ll 1$.
The bold lines correspond to the observational 
constraints (95 \%~{\rm CL}) on each model 
derived from the joint data analysis of 
WMAP 7yr, BAO, and HST. We also plot the boundary 
coming from $f_{\rm NL}^{\rm equil}>-214$ as well as the curves
corresponding to $f_{\rm NL}^{\rm equil}=-50$ 
and $f_{\rm NL}^{\rm equil}=-80$.}
\end{figure}

We study the following more general models
\begin{equation}
g(Y)=c_0+\sum_{p \neq 0} c_p Y^p\,,
\end{equation}
where $c_p$ are constants.
The power $p$ can be integer or some real number.
{}From Eq.~(\ref{ratioapp}) the ratio 
$c_s^2/\epsilon$ is given by 
\begin{equation}
\frac{c_s^2}{\epsilon} \simeq 
\frac16 \frac{\sum pc_p Y_0^{p-1}}
{\sum p(p+1) c_p Y_0^{p-1}}\,.
\label{ratiopower}
\end{equation}
For the single power $p$, i.e. 
$g(Y) = c_0 + c_p Y^p$, Eq.~(\ref{ratiopower}) 
reduces to 
\begin{equation}
\frac{c_s^2}{\epsilon} \simeq 
\frac{1}{6(p+1)}\,.
\label{ratiopower2}
\end{equation}
The conditions (\ref{dgy1}) and (\ref{dgy2}) 
give $pc_p>0$ and $p(p+1)c_p>0$, 
respectively, which demand that $p>-1$.
More precisely we require $c_p<0$ for $-1<p<0$
and $c_p>0$ for $p>0$.
In Fig.~\ref{fig5} we plot the line (\ref{ratiopower2}) in the $(\epsilon, c_s^2)$ 
plane for five different values of $p$ ($=-0.5, 0.5, 1, 2, 3$).
The ghost condensate model corresponds to $p=1$
with the tangent $c_s^2/\epsilon=1/12$.

We carry out the CMB likelihood analysis for the models
$g(Y)=c_0+c_p Y^p$ with $p=-0.5, 0.5, 1, 2, 3$
by employing the linear expansion given above.
The observational constraints shown in Fig.~\ref{fig5} 
with the bold lines
are derived by using the theoretical values of 
$n_{\rm s}$, $r$, and $n_{\rm t}$
given in Eqs.~(\ref{obser1}) and (\ref{obser2})
with the relation (\ref{ratiopower2}).
In the $(\epsilon, c_s^2)$ plane we also plot the border
corresponding to the WMAP 7yr lower bound 
$f_{\rm NL}^{\rm equil}=-214$.
The region above this border satisfies the observational 
constraint of the non-Gaussianity.
{}From Fig.~\ref{fig5} we find that there 
is no viable parameter space for $p \ge 3$
satisfying all the observational constraints. 
As long as the bold lines plotted in Fig.~\ref{fig5}
are above the border corresponding to 
$f_{\rm NL}^{\rm equil}=-214$, 
the models with $p<3$ can be compatible 
with the observational data.
If the future observations can place the bound on  
$f_{\rm NL}^{\rm equil}$ larger than $-80$, 
the models with $p>1/2$ can be ruled out
(see Fig.~\ref{fig5}).

Let us also discuss the case in which the function $g(Y)$ is 
the sum of different powers of $p$.
For example we consider the model
\begin{equation}
g(Y)=c_0+c_{1}Y+c_{-1}Y^{-1}\,.
\label{gycom}
\end{equation}
This corresponds to the dilatonic ghost condensate 
in the presence of the exponential potential, i.e.
$P=c_0 X+c_1 e^{\lambda \phi} X^2+c_{-1}e^{-\lambda \phi}$.
Substituting Eq.~(\ref{gycom}) into Eq.~(\ref{gY0}), we obtain
$Y_0=-c_0/(2c_1)$.
The conditions (\ref{dgy1}) and (\ref{dgy2}) translate into 
$c_1 (1-4c_1 c_{-1}/c_0^2)>0$ and $c_1>0$, respectively.
Since $Y_0=-c_0/(2c_1)>0$, we require that 
\begin{equation}
c_0<0\,,\quad c_1>0\,,\quad
4c_1 c_{-1}/c_0^2<1\,.
\end{equation}
{}From Eq.~(\ref{ratiopower}) we have 
\begin{equation}
\frac{c_s^2}{\epsilon}=\frac{1}{12}
\left( 1-\frac{4c_1c_{-1}}{c_0^2} \right)\,.
\label{csline}
\end{equation}
If $c_{-1}>0$, then the tangent of the line 
(\ref{csline}) gets smaller relative to that 
in the ghost condensate model.
Figure \ref{fig5} shows that the allowed parameter
space tends to be narrower for smaller $c_s^2/\epsilon$. 
The existence of a viable parameter demands the 
following condition
\begin{equation}
\frac{c_1 c_{-1}}{c_0^2} \lesssim 0.1\,.
\label{conf}
\end{equation}
The effect of the negative exponential potential 
$V=-c_{-1}e^{-\lambda\phi}$ (with $c_{-1}>0$) 
needs to be suppressed to be consistent 
with the bound (\ref{conf}).
In contrast, the tangent of the line (\ref{csline}) 
gets larger than $1/12$ when $c_{-1}<0$.
The effect of the positive exponential potential 
$V=-c_{-1}e^{-\lambda \phi}$ (with $c_{-1}<0$) 
makes it easier to satisfy the observational constraints.

\subsection{Class (ii)}

In the DBI model, inflation occurs in the ultra-relativistic regime
($Y \approx 1/2$) under the condition (\ref{ccon}).
In this case the denominator of $\epsilon$ 
in Eq.~(\ref{epDBI}) is much larger than its numerator.
Since the linear expansion around $Y=1/2$ is not possible
in such cases, we need to treat this class of models separately.
Let us take the function of the form
\begin{equation}
g(Y)=-\frac{c}{Y} \left[ 1+f(Y) \right]\,,
\label{gYclassII}
\end{equation}
in which case Eqs.~(\ref{epsilon}) and (\ref{cs2}) give
\begin{eqnarray}
& & \epsilon=-\frac{3Yf'(Y)}{1+f(Y)-2Yf'(Y)}\,,
\label{epsilon2} \\
& & c_s^2=\frac{f'(Y)}{f'(Y)+2Yf''(Y)}\,.
\label{cs22}
\end{eqnarray}
In the DBI model with $f(Y)=\sqrt{1-2Y}/c$, we can realize inflation 
in the regime $|f'(Y)|=1/[|c|\sqrt{1-2Y}] \ll 1$ and 
$f(Y) \ll 1$ with $Y \approx 1/2$, so that $\epsilon \ll 1$.

\begin{figure*}
\includegraphics[height=3.0in,width=7.0in]{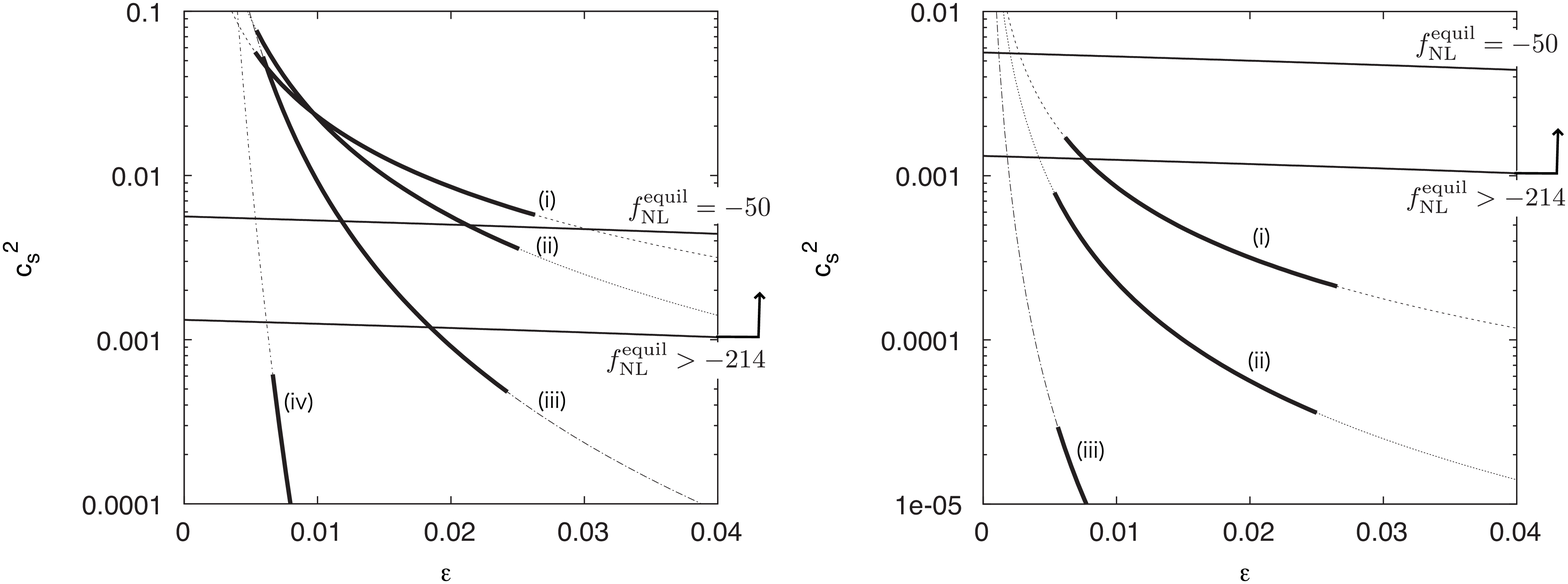}
\caption{\label{fig6}
The thin curves show the relation between $\epsilon$ and $c_s^2$ 
for the models (\ref{gYclassII}) with (\ref{fYII}).
The left and right panels correspond to $c=10^3$ and $c=10^4$, 
respectively, with (i) $m=0.3$, (ii) $m=0.5$, (iii) $m=0.7$, 
and (iv) $m=0.9$.
The bold curves represent the observational constraints (95 \%~{\rm CL}) 
derived from the CMB likelihood analysis in terms of $n_{\rm s}$, 
$r$, and $n_{\rm t}$.
We also plot the borders corresponding to the WMAP 7yr bound 
$f_{\rm NL}^{\rm equil}>-214$ as well as 
$f_{\rm NL}^{\rm equil}=-50$.}
\end{figure*}

We study the models (\ref{gYclassII}) with 
\begin{equation}
f(Y)=(1-2Y)^m/c\,.
\label{fYII}
\end{equation}
{}From Eqs.~(\ref{epsilon2}) and (\ref{cs22}) we have
\begin{eqnarray}
& & \epsilon=\frac{6mY}{c(1-2Y)^{1-m}+1+2(2m-1)Y}\,,\\
& & c_s^2=\frac{1-2Y}{1-2(2m-1)Y}\,.
\end{eqnarray}
We consider the case in which inflation occurs for the values of
$Y$ slightly smaller than $1/2$, 
while satisfying the condition $c(1-2Y)^{1-m} \gg 1$.
Since we require $\epsilon \simeq 3m/[c(1-2Y)^{1-m}]>0$
and $c_s^2 \simeq (1-2Y)/[2(1-m)]>0$, we have
either $0<m<1$ with $c>0$ or $m<0$ with $c<0$.
The following relation also holds between 
$c_s$ and $\epsilon$:
\begin{equation}
c_s^{2(1-m)} \simeq \frac{3m}{c[2(1-m)]^{1-m}}
\frac{1}{\epsilon}\,.
\label{curve}
\end{equation}

In Fig.~\ref{fig6} we plot the curve (\ref{curve}) in the 
$(\epsilon, c_s^2)$ plane 
for four different values of $m$ ($=0.3, 0.5, 0.7, 0.9$).
The left and right panels correspond to the cases
$c=10^3$ and $c=10^4$, respectively.
We also show the observational bounds constrained by 
$n_{\rm s}$, $r$, and $n_{\rm t}$ (plotted as the bold lines) 
as well as the curves corresponding to $f_{\rm NL}^{\rm equil}=-214$
and $f_{\rm NL}^{\rm equil}=-50$.

When $c=10^3$ there exists some allowed parameter space for the models 
with $m \le 0.7$ (including the DBI model with $m=0.5$), 
but for $m \ge 0.9$ the WMAP 7yr bound of the non-Gaussianity 
excludes the parameter region constrained by the linear perturbations.
For larger $c$ the theoretical curves in Fig.~\ref{fig6} shift to 
the regions with smaller $c_s$, so that the constraint from 
the non-Gaussianity tends to be more important.
For $c=10^4$ the right panel of Fig.~\ref{fig6} shows that, 
the models with $m \ge 0.5$ do not have the viable parameter space 
satisfying all the current observational constraints.
If the future observations can reach the level of the lower limit of
the non-Gaussianity with $|f_{\rm NL}^{\rm equil}|={\cal O}(10)$, 
then it is possible to place tighter constraints further
(see the curves in Fig.~\ref{fig6} corresponding to $f_{\rm NL}^{\rm equil}=-50$).

\section{Conclusions}
\label{Conclusions}

We have studied the observational constraints on assisted k-inflation 
models in which the multiple scalar fields join an effective single-field
attractor described by the Lagrangian $P=X g(Y)$ with 
$Y=Xe^{\lambda \phi}$. The canonical field with the exponential 
potential, $P=X-ce^{-\lambda \phi}$ (i.e. $g(Y)=1-c/Y$), 
is one of the simplest examples giving rise to assisted inflation.
The effective slope $\lambda$ along the inflationary attractor
is given by $\lambda=\left( \sum_{i=1} 1/\lambda_i^2 
\right)^{-1/2}$, which is smaller than the slopes $\lambda_i$
for each exponential potential.
The same structure holds for the k-inflation models 
with the Lagrangian $P=X g(Y)$ for arbitrary
functions of $g(Y)$.

Along the effective single-field attractor, the inflationary observables 
are in general given by Eqs.~(\ref{obser1})-(\ref{obser3}).
In Sec.~\ref{constraint} we have confronted four models of assisted 
inflation with the recent observations of CMB combined with 
BAO and HST. For the canonical field with the exponential 
potential the effective slope $\lambda$ is constrained to be 
$0.086<\lambda<0.228$.
The tachyon field needs to have a small kinetic energy relative
to its potential energy for the realization of inflation, in which case
the observational bound on the variable $Y$ is given 
by Eq.~(\ref{Ycon}). Since the field propagation speed
$c_s$ is close to 1 in this case, the primordial non-Gaussianity
remains to be small for the tachyon model.

In the dilatonic ghost condensate model
the non-Gaussianity provides additional constraints to 
those derived by the spectra of scalar and tensor perturbations.
As we see in Fig.~\ref{fig2}, the WMAP 7 yr limit 
$f_{\rm NL}^{\rm equil}>-214$ reduces the allowed 
parameter space of the parameter $\lambda$.
If the lower bound on $f_{\rm NL}^{\rm equil}$ reaches
the level of $-100$ in future observations, it will be possible 
to rule out the dilatonic ghost condensate model.
In the DBI model the level of the non-Gaussianity depends
on the field propagation speed $c_s$ as well as
the constant $c$ associated with the energy scale of 
the potential. For larger $c$, $|f_{\rm NL}^{\rm equil}|$
tends to increase, so that the models can be constrained 
by the additional information coming from the non-Gaussianity.
In fact the DBI model with $c \gtrsim 10^4$ is 
excluded by the WMAP 7yr data.

We have extended the analysis to more general functions 
$g(Y)$ by classifying the assisted k-inflation models 
into two classes. The first class consists of the models in which 
inflation occurs around $Y=Y_0$ satisfying the condition 
$g(Y_0)+Y_0g'(Y_0)=0$. The representative models of 
this class are $g(Y)=c_0+\sum_{p \neq 0}c_p Y^p$, 
which includes the dilatonic ghost condensate. 
{}From the CMB likelihood analysis combined with the 
non-Gaussianity bound we have found that the single-power
models $g(Y)=c_0+c_p Y^p$ with $p \ge 3$ are 
ruled out. The second class consists of the models with 
the speed limit of the field, which includes 
the DBI model as a specific case.
We have carried out the CMB likelihood analysis for the 
functions $g(Y)=-(c/Y)[1+(1-2Y)^m/c]$ ($m<1$) and
showed that the models with larger $m$ and $c$
tend to be observationally disfavored by taking into account 
the non-Gaussianity bound.

In this paper we have evaluated the inflationary observables
under the assumption that the solutions are on the assisted attractor
described by the effective single field.
In order to end inflation the solutions need to exit from this regime. 
This can be achieved by treating the validity of
our Lagrangian $P=Xg(Y)$ only within some limited range of 
field values. With some suitable modification of the Lagrangian 
it is possible to lead to the graceful exit of inflation \cite{kinflation}.
Another possibility is that k-inflation ends with a phase transition 
as in hybrid inflation \cite{Arkani2}.
It will be also of interest to study the case where the observed
CMB anisotropies correspond to the epoch before the multiple 
fields join the inflationary attractor.
In this case the trajectory in field space is curved, so that   
isocurvature perturbations can contribute to 
adiabatic perturbations \cite{Bruce}.
We leave these issues for future work.

\section*{ACKNOWLEDGEMENTS}
We thank Antonio De Felice for his help to run the Cosmo-MC
code and for useful discussions.
S.\,T.\ thanks financial support for JSPS (No.~30318802)
and the Grant-in-Aid for Scientific Research 
on Innovative Areas (No.~21111006). 

\end{document}